# ALDAS: Audio-Linguistic Data Augmentation for Spoofed Audio Detection

Zahra Khanjani
Information Systems
UMBC

Christine Mallinson
Language, Literacy, and Culture
UMBC

James Foulds
Information Systems
UMBC

Vandana P Janeja
Information Systems
UMBC

**Abstract:** Spoofed audio, i.e. audio that is manipulated or AI-generated deepfake audio, is difficult to detect when only using acoustic features. Some recent innovative work involving AI-spoofed audio detection models augmented with phonetic and phonological features of spoken English, manually annotated by experts, led to improved model performance. While this augmented model produced substantial improvements over traditional acoustic features based models, a scalability challenge motivates inquiry into auto labeling of features. In this paper we propose an AI framework, Audio-Linguistic Data Augmentation for Spoofed audio detection (ALDAS), for auto labeling linguistic features. ALDAS is trained on linguistic features selected and extracted by sociolinguistics experts; these auto labeled features are used to evaluate the quality of ALDAS predictions. Findings indicate that while the detection enhancement is not as substantial as when involving the pure ground truth linguistic features, there is improvement in performance while achieving auto labeling. Labels generated by ALDAS are also validated by the sociolinguistics experts.

## 1. Introduction

With advances in AI, it is possible to rapidly and easily create highly realistic fake content [1]. As fake media becomes pervasive, it becomes more challenging for the public to discern fake and real content. Fake audio poses a particular threat in its ability to disseminate misinformation and perpetrate fraudulent activities. For example, in 2019 AI-based software was used to mimic a CEO's voice and swindle over $243,000 by phone [2]. Robocalls have also perpetrated confusion and misinformation about political elections via audio alone [3]. Spoofed audio, both human-generated (e.g. mimicry and replay attacks) and AI-generated (e.g. audio deepfakes), thus represent a clear social threat in the era of online communication [4]. Yet, audio deepfake detection, compared to other fake media (especially image and video) detection, remains understudied [1].

Most research methods for spoofed audio detection (SAD) rely on common labeled datasets and algorithm-driven approaches. Most studies extract audio features--such as waveforms, Linear Frequency Cepstral Coefficients (LFCCs), Mel Frequency Cepstral Coefficients (MFCCs) or spectrograms--and then train a deep learning binary classification model using these features. Khochare et al. [5], used Temporal Convolutional Network (TCN) and Spatial Transformer Network (STN) to classify a benchmark Fake or Real dataset [6] using MEL spectrogram as the input feature of the audio data. Their results showed that TCN worked better than the other models on the FoR dataset. However, being limited to the FoR dataset reduces the generalizability of the model, since it only consists of one sub-type of audio deepfake called text-to-speech.

Newer studies are adopting innovative approaches to detect fake audio. Some studies use deep neural networks applied to acoustic features, as in [7, 8, 9]. Blue et al. apply knowledge from articulatory phonetics and fluid dynamics to identify fake audio, determining whether a given speech sample could have been produced by a human vocal tract [10]. In another study, Li et al. [11] compare perceptual features with regular acoustic features such as Mel Frequency Cepstral Coefficients (MFCCs). When applied on the Audio Deep synthesis Detection challenge (ADD) 2022 dataset [12], results show that even a simple perceptual feature extraction can significantly affect detection performance. However, replicating this work does not seem possible, as the challenge [12] does not provide the public with the dataset after the challenge timeline.

Such studies demonstrate the utility of applying an analysis of linguistic properties of sound signals to the audio deepfake detection challenge, but this area of inquiry can be deepened. Recent work [4] demonstrated how manually extracted linguistic features can improve one of the common baselines of Automatic Speaker Verification and Spoofing Countermeasures Challenge (ASVspoof 2021) [13]. The authors incorporated expert knowledge to extract what were called Expert Defined Linguistic Features (EDLFs). The EDLFs augmented the training data, which was used to train different machine learning models for the purpose of SAD. Findings showed that Logistic Regression classifier augmented by EDLFs outperformed one the best performing baselines, a pre-trained model of Light Convolutional Neural Network (LCNN) followed by Long Short-Term Memory (LSTM) [14, 15] and input with Linear Frequency

Cepstral Coefficients (LFCCs). The EDLFs were derived from linguists' perceptual annotations of how a set of commonly occurring features of spoken English are realized in a given set of real and fake audio clips. Four of the EDLFs used were commonly occurring phonetic and phonological features: pitch, pause, word-initial and word-final release bursts of English consonant stops, and audible intake or outtake of breath. Experts in sociolinguistics listened for the presence or absence of each feature and any anomalies in their expected production. They also assessed overall audio quality, including listening for disturbance or distortion to the sound signal (e.g. audio that was unusually tinny, robotic, or compressed). Audio quality is not typically included in sound quality assessment in the literature; see [16], which describes background noise as an anomaly/low quality. For more information regarding the expert annotation process for EDLFs, see the primary paper [4].

One challenge with the EDLF-based approach is the manual annotation process, which limits the size of training data. This challenge motivates a need to explore auto labeling approaches using AI models that are still accompanied by domain expert input for the purposes of model training and validation. Encoding breath for SAD has been addressed by some other studies, such as Breathing Talking Silence encoder [24] and [25]; however, in those studies, linguists were not involved, so potential extension with regard to additional analysis of this or other linguistic features is generally not explored.

In this paper, we automated the manual process for the labeling of these EDLFs from audio data. We call this automated process ALDAS: Audio-Linguistic Data Augmentation, and it is designed to use AI models for labeling data with linguistic features for better spoofed audio detection. To our knowledge, no existing work has examined auto labeling of audio samples with linguistic features that have been validated by experts with the goal of improved SAD baselines.

This paper is organized as follows: Sections 2 and 3 describe the methodology and validation respectively, and Section 4 discusses the experimental setting and results. Finally, Section 5 discusses the conclusion and future directions.

Our overall methodology is depicted in Figure 1. We use the pre-trained VGGish [17] model to extract representations of audio data. The audio data representations are then input to the Convolutional Neural Network (CNN) classifier, in addition to training labels for the EDLFs that were identified by the sociolinguistics experts. The CNN classifier has been fine-tuned with respect to the EDLFs. Due to the small sample size, cross validation is used for tuning the hyper-parameters of the CNN classifier. The output of the classifier is the predicted Expert Defined Linguistic Features ($EDLF^p$), which is used for SAD through an ensemble model with baselines for performance evaluation. For Step 1 of ALDAS, we auto-label the EDLFs in the EDLF-prediction task. Not all of the EDLFs can be auto-labeled using ALDAS, as described below.

**Which EDLFs?** ALDAS is used for auto labeling three EDLFs: audible intake or outtake of breath (labeling presence or absence), pitch (labeling expected or anomalous production), and audio quality (labeling expected or anomalous production). While the anomalous pause feature is excluded to avoid overfitting, anomalous consonant burst is removed due to causal-based feature importance, as reported in another work [23]. The results of [23] show the most effective EDLF is anomalous audio quality, followed by anomalous pitch, anomalous pause, and presence of breath, respectively.

*Definition 1:*
Given an audio clip $a_i$, with the linguistic feature set (EDLFs) where:
$$a_i^{EDLFs} = [\, a_i^{\text{presence-of-breath}},\; a_i^{\text{pitch-anomaly}},\; a_i^{\text{audio-quality-anomaly}}\,]. \qquad (1)$$

The distributions of the EDLFs can be imbalanced, as was the case in our training dataset. To address imbalanced data for auto-labeling pitch and breath, we over-sampled the minor class via the SMOTE method [18].

For auto labeling the three EDLFs, we applied VGGish, a transfer learning model based on Convolutional Neural Networks (CNNs) trained on massive audio data.

## 2. ALDAS Methodology

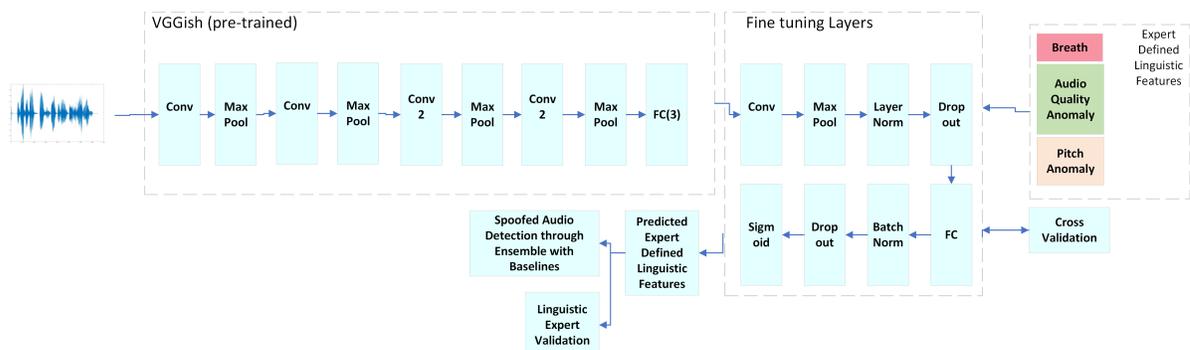

Figure 1- ALDAS Overall Methodology. The two boxes "VGGish (pre-trained)" and "Fine tuning Layers" are called the EDLF-prediction task. FC means Fully Connected layers.

**VGGish (The Front End):** VGGish, released by Google AI [17] is an audio adaption of VGG (Visual Geometry Group) network architecture that usually is used in image classification [19].

VGGish is pre-trained on millions of YouTube videos and has shown superior performance when used as audio feature representations in different tasks. We use VGGish representations as the input of the back end model for the linguistic feature auto labeling.

**Fine-tuning layers (The Back End):** In this phase, classifiers are trained with the VGGish representations and the expert labeled data. The dataset contains 840 audio samples, for which the EDLFs are labeled (supervised learning). Among different classifiers used (Support Vector Machine, Logistic Regression, Multi-layer Perceptron, and CNN), the CNN classifier had outstanding performance, shown in Figure 1. We used multiple regularization techniques including dropout, layer normalization, and batch normalization, per Figure 1, to overcome the problem of small sample size. To achieve the final settings, cross validation k (the number of folds) = 5 is used.

**ALDAS:** ALDAS is depicted in Algorithm 1. ALDAS is trained on the feature labels provided by the sociolinguistics experts; therefore, the trained EDLF-prediction model can be used for any unseen dataset in the future. The auto-labeled EDLFs, which we now call predicted EDLFs or $EDLF^p$, are used as input features to a multi-layer perceptron (MLP) classifier to detect spoofed audio. The details of the best performing MLP model are: 'alpha'=0.01, 'early-stopping' = 'True', 'hidden-layer-sizes'= (32, 16), 'solver': 'adam'. The results from MLP are then provided as input to an ensemble model with the pre-defined baseline, in the last step of ALDAS, as seen in the Algorithm 1. The baseline can be anything through GetBaselineOutput function in Algorithm 1, we considered common baselines of ASVspoof 2021 [13] to test the performance of ALDAS in SAD.

---
**Algorithm 1** Audio-Linguistic Data Augmentation for Spoofed audio detection
---
**Require** A mini-batch speech data $S = s_{genuine}, s_{spoofed}$.
A mini-batch speech VGGish representations $V = v_{genuine(VGGish)}, v_{spoofed(VGGish)}$.
**Initialize** weights for auto-labeling and baseline outputs
**Initialize** lists to store auto-labeling and baseline outputs
**If** $EDLF^p = presence - of - breath$ or $pitch - anomaly$:
  $V = SMOTE(V)$,
**for** number of training iterations **do**
  **for** k-th mini-batch **do**
    **for** each audio_sample_ in V **do**
      Pass through EDLF-prediction
**return** $EDLF^p$
**Obtain baseline output:**
  baseline_output = GetBaselineOutput$((s_i \in S))$
  baseline_outputs.append(baseline_output)
**Spoof Detection with** $EDLF^p$:
  SpoofDet_output = MLP($EDLF^p$)
  SpoofDet_outputs.append(SpoofDet_output)
**Ensemble of the outputs:**
  classification_result = [weight_ALDAS $\times$ (SpoofDet_outputs)|weight_baseline $\times$ (baseline_outputs)]
**return** predicted labels(spoofed, genuine)

---

$ALDAS_h$ **vs** $ALDAS_s$: Given an audio clip $a_i$ and ALDAS Algorithm 1, soft and hard thresholding of ALDAS is described as follows:

$$ALDAS^h(a_i) \implies EDLF^p \; where \; label \in 0,1 \quad (2)$$

$$ALDAS^s(a_i) \implies EDLF^p \; where \; label \in p_{ij} \quad (3)$$

where $p_{ij}$ is the probability that audio clip $a_i$ has a $EDLF_j$. Soft and hard analysis is provided to find the final strategy for auto-labeling in ALDAS.
experts then listened to and labeled the audio clips

## 3. ALDAS Validation
The main components of our validation process are: a) SAD metrics when $EDLF^p$ from ALDAS are used to augment the data for the common baselines through the ensemble models (the first test/ unseen set, which was 15% of the whole primary dataset), and b) $EDLF^p$ is validated by the sociolinguistics experts (the second test/unseen set). We explain the first one a) in (SAD Metrics), and b) in the subsection (Expert Validation).

**Spoofed Audio Detection Metrics:** Since our dataset is balanced by number of spoofed versus genuine samples, accuracy can be used as an evaluation metric for SAD. Other classification metrics are also utilized, such as ROC AUC, harmonic mean of precision and recall (F1 score), as well as Equal Error Rate (EER), a very common metric in SAD.

**Expert Validation:** The auto-labeled $EDLF^p$ were applied to a dataset unheard by the sociolinguistics experts. For this dataset, the final labels (genuine and spoofed) were also hidden from the experts. We show ROC AUC based on each $EDLF^p$ as validated by experts.

## 4. Experimental Results and Settings
### 4.1. Datasets:
We used two datasets for our experimentation.

**The Primary Dataset:** Although several datasets are available for SAD, to the best of our knowledge none of them contain all types of spoofed audio attacks, including AI generated (text-to-speech, or TTS, and voice conversion, or VC) and non-AI generated audio data (mimicry and replay attacks). Therefore, we created a data set containing samples from the common datasets [13, 20, 21] in addition to new spoofed audio samples. The new audio data we added include mimicry[1], some TTS and VC samples generated by available online tools[2], and popular generative networks such as [22]. The number of the audio data is 840, almost balanced in terms of spoofed versus genuine samples. For spoofed samples: 23%, 30%, 31% and 18% are replay attack, TTS, VC, and mimicry, respectively. All of these were examined by sociolinguistics experts and labeled with EDLFs. From the whole dataset, 15% is completely held out from the modeling phases as our first unseen/test set for which the types of attacks distributions are preserved, and the test set is also balanced in terms of genuine versus spoofed samples. The average duration of the audio clips is 4.01 seconds. All clips are in English.

---
[1] Extracted from YouTube channels mimicking celebrities and political figures
[2] https://www.resemble.ai/

**The Linguistic Expert Validation Dataset:** For the linguistic expert-validation dataset we randomly selected 100 audio clips from the evaluation set of ASVspoof 2021 [13] dataset, Deepfake (DF) task. This is considered our second test dataset, which is also previously unheard by the sociolinguistics experts, and used for evaluation of the auto labeling task.

### 4.2. Results:

We now discuss the experimental settings, rationale, results obtained for SAD, and evaluation of auto labeling.

**Results in Spoofed Audio Detection:** The impact of linguistic augmentation is demonstrated on the ASVspoof common baselines (LFCC-Gaussian Mixture Model (GMM), LFCC-LCNN and RawNet2) as shown in Figure 2. As it indicates, the best performing model is an ensemble of a logistic regression model input with true EDLFs, and LFCC-LCNN (EDLF-LR|LFCC-LCNN). The second best is (EDLF-LR|RawNet2). ALDAS ensemble with any of the baselines, has the third best performance which means an ensemble model of a multi-layer perceptron augmented by $EDLF^p$, and the baseline. These results indicate how augmenting with the the predicted linguistic features improves the baselines. The increase in the performance is not as much as when true EDLFs are involved, but it still has significant impact to show the value of auto labeling EDLFs.

Additionally, the final $EDLF^p$ is obtained from $ALDAS^s$ and showed better performance on the validation sets compared to $ALDAS^h$.

Also, Figure 3 shows the details of the metrics for a common baseline (pre-trained LCNN followed by LSTM input with LFCCs, for which more details are presented in [13]) and when ALDAS is applied. The best performing weights to be used in the ensemble model are obtained across multiple iterations of the Algorithm 1, and finally (Weight$_{ALDAS}$ = 0.6) is chosen. As shown, ALDAS has improved almost all of the metrics across the training and testing datasets. Additionally, Table 1 indicates how both manual and predicted EDLFs reduce Equal Error Rate on unseen data.

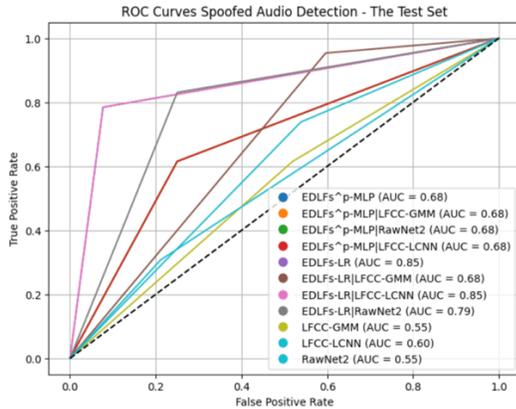

Figure 3- ROC AUC for the test set. The true EDLF model, which is a logistic regression input with EDLFs, ensemble with LFCC-LCNN baseline, has the best performance

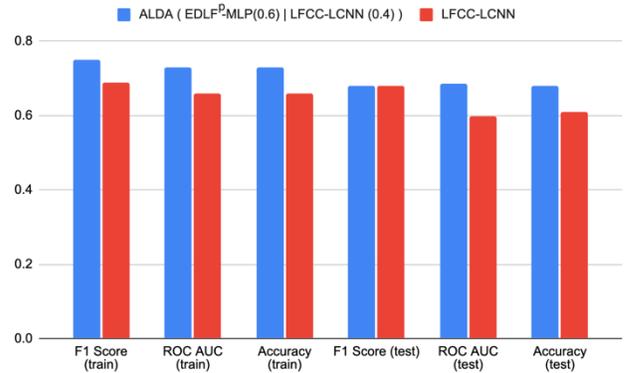

Figure 4- LFCC-LCNN baseline and ALDAS metrics

TABLE I
EER FOR THE BEST BASELINE ALONE, AND WHEN TRUE EDLFs AND PREDICTED EDLFs ARE INVOLVED; FOR THE TEST SET

| Dataset | LFCC-LCNN | $EDLF\text{-}LR$|LFCC-LCNN | $EDLF^p\text{-}MLP$|LFCC-LCNN |
|---|---|---|---|
| Train | 0.33 | 0.145 | 0.25 |
| Test | 0.39 | 0.14 | 0.31 |

### 4.3. Linguistic Expert Validation

The Linguistic Expert Validation data was not previously heard by the sociolinguistics experts; following the ALDAS auto labeling on this data, the experts then listened to and labeled the audio clips for accuracy computation of the $EDLF^p$. This dataset plays the role of an additional test set to validate the evaluation results. For this dataset, ALDAS was used to label the three identified EDLFs for presence or absence (breath) and expected or anomalous production (pitch, audio quality). Among all 300 auto labels for $EDLF^p$, the accuracy when considering a threshold of 0.5 is 78%; (Overall accuracy to emulate real world cases where the accuracy is important across all of the EDLFs, is presented as a single accuracy). Figure 4 shows ROC AUC scores for presence of breath, anomalous audio quality, and anomalous pitch. These results reveal that the performance of ALDAS for labeling features is the most promising for the feature of breath.

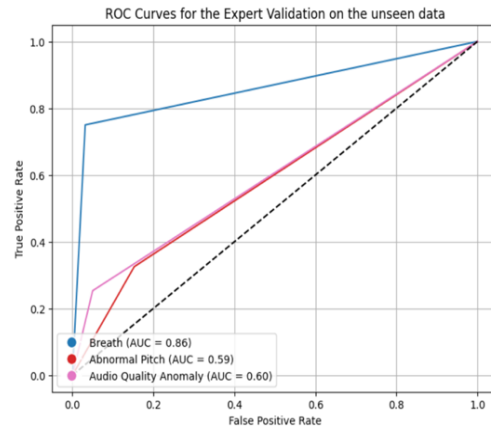

Figure 2- ROC AUC for the Linguistic Expert Validation on the

All of the auto labeling models are trained on CPU and incur no computational expenses; the only item for which we used GPU[3] is applying the baselines on the dataset. Also, the total parameters of each auto-labeling models are 1442305, 721217, and 14817 for presence of breath, pitch and audio quality anomalies respectively. For all of the auto-labeling models the optimizer is Adam, and the loss is binary-crossentropy.

## 6. CONCLUSION AND FUTURE WORK

This work has introduced ALDAS: Audio-Linguistic Data Augmentation for Spoofed audio detection. ALDAS is a CNN-based model trained on audio samples labeled with several features of spoken English that were also annotated and validated by sociolinguistics experts.

Although ALDAS does not exceed pure expert annotated features, it establishes a mechanism of auto labeling linguistic features and their viability for SAD. Moreover, the auto labeled features from ALDAS significantly improve the common baselines of ASVspoof 2021. These baselines could not show good performance when faced with a dataset containing mixed types of attacks. In future work, we plan to extend ALDAS to more accurate predictions of these and other linguistic features. Through this study we also demonstrate the importance of developing these models in cooperation with and under the supervision of linguists and domain experts who are able to inform auto-annotation models and validate results. These types of features account for the uniqueness of variation in human language use, which is largely absent from current detection models. Adding ALDAS to neural-network-based representations of audio will be considered as another fruitful avenue in SAD.

## 7. ACKNOWLEDGMENTS

Authors would like to acknowledge support from the National Science Foundation Awards #2210011 and #2346473. All codes and audio samples are available through our GitHub repository [26]. Authors would like to acknowledge the contributions of Lavon Davis, who assisted with the linguistic labeling of 500 audio samples, as well as Noshaba Nasir Bhalli and Chloe Evered for their assistance with experiments and data analysis.

---

[3] Free version of Google Colab GPU